\documentstyle[amsmath,amssymb,aps,epsf,eqsecnum,graphicx]{revtex}

\begin{document}
\draft

\title{ Relativity of pure states entanglement }
\author{Karol {\.Z}yczkowski$^{1}$\footnote{on leave from
Instytut Fizyki,  Uniwersytet Jagiello{\'n}ski,
 ul. Reymonta 4, 30-059 Krak{\'o}w, Poland}
 and Ingemar Bengtsson$^2$}
\address{$^1$Centrum Fizyki Teoretycznej, Polska
  Akademia Nauk, \\ Al. Lotnik{\'o}w 32/46, 02-668 Warszawa, Poland}
\address{$^2$Stockholm University, SCFAB, Fysikum \\
 S106 91 Stockholm, Sweden}
\date{ August 6,  2001}
\maketitle

\begin{abstract}
Entanglement of any pure state of an $N \times N$  bi-partite quantum
system may be characterized by the vector of coefficients arising by
its Schmidt decomposition.
 We analyze various measures of entanglement
derived from the generalized entropies of the vector
of Schmidt coefficients.
 For $N \ge 3$ they generate different
ordering in the set of pure states and for some states
their ordering depends on the measure of entanglement used.
This odd-looking property is acceptable,
 since these incomparable states
cannot be transformed to each other with unit efficiency by any local
operation.  In analogy to special relativity the set of pure states
equivalent under local unitaries has a causal structure so that at each
point the set splits into three
parts: the 'Future', the 'Past' and the set of noncomparable states.

\end{abstract}

\section{Introduction}

Due to growing interest in quantum information
the properties of quantum entanglement
became a subject of intensive research. Entanglement gives rise to 
the peculiarly quantum mechanical correlations that may exist between 
two subsystems. In spite of several papers published yearly on this subject
we do not know the answers to even some very basic questions in this field. 
An example of such a question: 'Is a given quantum state
(pure or mixed) of a composite system separable or entangled?'
For the simplest case of $2 \times 2 $ and $3 \times 2$ systems
this problem is solved by the Peres-Horodeccy criterion
\cite{Pe97,HHH}, which offers a simple
condition sufficient and necessary for separability.
On the other hand, in the general case of a $K \times M$ bi-partite
system such a constructive condition remains still unknown \cite{primer,HHH01}.

Another important issue is the question: 'How much is a given quantum 
state entangled?'. In other words we are looking for a function of the 
states that quantifies entanglement. Since we do not quite understand what 
entanglement means this is a difficult question, but what we can do 
is to write down conditions that such a measure has to satisfy. A key 
condition is that the expected entanglement cannot increase unless the two 
subsystems interact in a quantum mechanical fashion. We therefore require 
monotonicity under local operations and classical communication (LOCC) 
\cite{BBPS96} \cite{BDSW96}. 
The idea is that we have two subsystems A and B and two experimenters 
Alice and Bob who are allowed to manipulate one subsystem each in any way 
they please. They are also allowed to tell each other, using classical 
communication, what unitary transformations and measurements they perform 
on their respective subsystems so that Alice can allow her manipulations 
to depend on what Bob does and what the results of his measurements are 
(and conversely for Bob). There exists a very beautiful characterization of 
LOCC operations due to Nielsen \cite{Ni99} which we will quote in due course. 

Now it is known \cite{Vi98} that monotonicity under LOCC operations does not 
suffice to single out a unique measure of entanglement. A major point 
of the present paper is precisely to study this non-uniqueness in detail. 
The situation changes if we consider the asymptotic regime, in which it 
is assumed that we have a source producing infinitely many copies of our 
quantum system. In this situation suitable continuity requirements can be 
added to the list of properties that an entanglement measure should satisfy, 
and there are uniqueness theorems \cite{DHR01} that show that on pure states 
the only measure that fullfills all the axioms must agree with a particular 
measure known as the entropy of entanglement when evaluated on pure states.  
In the present paper we stay in the regime of one copy however. 

Classifying measures of entanglement and selecting the physically relevant 
ones is a subject of vivid recent interest \cite{Vi98,DHR01,Ru00,Ho01}. One 
line of research starts with the investigation of invariants of local 
unitary transformations, like Schmidt coefficients for pure states \cite{Pe93}. 
Several such invariants for mixed states of a bi--partite system were 
recently found \cite{LPS99,GRB98,EM99,Ma00,CS00,Lo01,KZ00},
and one may try to construct
existing measures of entanglement out of them or define
new measures with attractive properties.
Another possible approach is to quantify the entanglement
by the distance of the given state to the closest separable state
\cite{VP98,WT98}, or to look for the best 
separable approximation of an entangled state \cite{LS98}. 

The aim of this paper is to elucidate some
geometrical  properties of the  entanglement of pure states
of a composite $N \times N$ system.
Using the concept of the {\sl Schmidt decomposition}
we illustrate the observation \cite{Vi98} that in the regime of one copy 
there exists infinitely many measures of entanglement
which generate different orderings of pure states. We establish a link
between different measures based on the generalized entropies
of the vector of Schmidt coefficients and the various distances
between closest separable states \cite{VP98}.

Applying Nielsen's characterization of LOCC, that is using the concept of 
{\sl majorization} \cite{Ni99,Vi99} (to be introduced below), we show that
any state $|\psi\rangle$ gives rise to a natural decomposition
of the set of all pure states into four sets:
a) the measure zero set of {\sl interconvertible} states
 which may be obtained from $|\psi\rangle$ by local unitary transformation,
b) the states accessible from $|\psi\rangle$ by
nonunitary local transformations,
c) the states which can be transformed locally into
$|\psi\rangle$,  and
d)---the set of incomparable states.
Once all pure states that can be transformed into each
other by local unitary transformations have been identified with each
other, the decomposition consists of only three sets, b), c) and d).
This picture mimics the well known structure of the light cone
in special relativity, which divides the space--time into
'Future', 'Past' and the set of space--like, incomparable events. 
Hence our title (see also \cite{HAHL00}). 
The same analogy works for the entirely different problem
of non--unitary evolution of mixed states under the
action of external random fields. Interestingly, in this case
the direction of the arrow of time is reversed.

The paper is organized as follows.
In section II we review the definitions of the Schmidt decomposition
of a pure state, quantum
R{\'e}nyi entropies and majorization, while in Section III
we discuss different measures of entanglement,
analyze nonhermitian evolution of the density matrices,
for which the majorization theory is applicable, and the corresponding
local transformations of pure states. 
Propositions emphasizing the geometrical interpretation of the
Weyl chamber consisting of ordered eigenvalues of a density matrix (or
ordered vector of the Schmidt coefficients for a pure state of a
bi-partite system) are proved in Appendix A.

\section{Mathematical tools and definitions}
\subsection{ Schmidt decomposition}

Consider a pure state $|\psi\rangle$
of a composite Hilbert space ${\cal H}={\cal H}_A \otimes {\cal H}_B$
of size $N^2$. Introducing an orthonormal basis $\{ |n\rangle
\}_{n=1}^N$
 in each subsystem, we may represent the state as

\begin{equation}
 |\psi\rangle = \sum_{n=1}^N
 \sum_{m=1}^N C_{mn} |n\rangle \otimes |m\rangle
.
\label{state}
 \end{equation}
The complex matrix of coefficients $C$ of size $N$ needs not to be
Hermitian nor normal. Its singular values
(i.e. square roots of
eigenvalues $\lambda_k$ of the positive matrix $C^{\dagger}C$)
determine the Schmidt decomposition \cite{Sc06,Pe93,EK95}
\begin{equation}
|\psi\rangle = \sum_{k=1}^{N} \sqrt{\lambda_{k}}
 |k{\prime}\rangle
\otimes |k^{\prime\prime}\rangle,
 \label{VSchmidt}
\end{equation}
where the basis in $\cal H$  is
transformed by a local unitary transformation $U \otimes V$. Thus
$|k{\prime}\rangle = U |k\rangle$, and
$|k{\prime\prime}\rangle = V
|k\rangle$,
 where $U$ and $V$ are the matrices of
eigenvectors of $C^{\dagger}C$ and $CC^{\dagger}$, respectively.
The remarkable thing is that only a single sum is involved. 
In the generic case of a non-degenerate vector $\vec \lambda$, the
Schmidt decomposition is unique up to two unitary diagonal matrices,
up to which the matrices of eigenvectors $U$ and $V$ are determined.
The normalization condition $\langle \psi|\psi\rangle=1$
 enforces $\sum_{k=1}^{N} \lambda_k=1$.
Thus the vector  ${\vec \Lambda}=(\lambda_1,...,\lambda_N)$
lives in the ($N-1$) dimensional simplex ${\cal S}_N$.
 Note that the Schmidt
coefficients $\lambda_k$ {\sl do not} depend on the initial basis
$|n\rangle \otimes |m\rangle$, in which the analyzed
state $|\psi\rangle$ is represented.
They are invariant with respect to any {\sl local  operations}
$U_A \otimes U_B$, and thus they may serve as ingredients
of any measure of entanglement. The Schmidt simplex is almost but not
quite the same as the space of orbits under local unitary
transformations; since
we can change the ordering of the eigenvalues by means of local
unitaries some further identification of points in the Schmidt simplex
has to be done before we have the space of orbits. This is discussed below
(section IIID).

The Schmidt coefficients of a pure state  $|\psi\rangle$
are equal to the eigenvalues of the
reduced density operator, obtained by partial tracing,
$\rho^A={\rm tr}_B(|\psi\rangle \langle \psi|)$.
The pure state is called {\sl entangled} if it
can not be represented in the product form
$|\psi\rangle=|\psi_A\rangle \otimes |\psi_B\rangle$,
where $|\psi_A\rangle \in {\cal H}_A$ and $|\psi_B\rangle
 \in {\cal H}_B$.
This is the case if and only if
there exists only one non-zero Schmidt coefficient, $\lambda_1=1$.

\subsection{Entangled mixed states}

Mixed states $\rho=\sum_{i} p_i |\varphi_i\rangle \langle \varphi_i|$
with $p_i>0$ and $\sum_i p_i=1$ will also be useful in our considerations.
A state $\rho$  is called a {\sl product state} if it can be represented
as a tensor product, $\rho_{prod}=\rho^A \otimes \rho^B$, where
$\rho^A$ acts in ${\cal H}_A$ and $\rho^B$ acts in ${\cal H}_B$.
A mixed state
$\rho$ is called {\sl separable} if it can be represented as a
convex combination of product states,
$\rho_{sep}=\sum_j q_j \rho^A_j \otimes \rho^B_j$, where
$q_j>0$ and $\sum_j q_j=1$. A mixed state which is not separable is
 called {\sl entangled}. It is easy to see
that for pure states both definitions are consistent.

\medskip

\subsection{Measures of entanglement}

There exist several different possibilities to quantify
quantum entanglement
-- for a review on this subject see e.g. \cite{HHH01,Ho01}.
Following Vedral and Plenio \cite{VP98} we assume that  each
measure of entanglement $E(\rho)$ has
 to satisfy the following conditions:

(E1) $E(\rho)=0$ if $\rho$ is separable,
(the condition 'if and only if' occurs to be too strong \cite{Ho01})

(E2) Local unitary
operations do not change the entanglement,
     i.e. $E(\rho)=E(U_A \otimes U_B \rho
         U_A^{\dagger} \otimes U_B^{\dagger})$

(E3) The expected entanglement cannot increase under operations involving local
measurements  and classical communication (LOCC), followed by post-selection:
\begin{equation}
\sum_{i=1}^M {\rm tr} \rho_i
   E\Bigl( \frac{\rho_i}{ {\rm tr} \rho_i} \Bigr)
\le E(\rho) ,
\label{Elocc}
\end{equation}
where $\rho_i=V_i\rho V_i^{\dagger}$, each of the operators $V_i$ is
local, i.e. $V_i=V_i^A\otimes V_i^B$,
 and the set of $M$ these operators defines an
positive operator valued measure (POVM) \cite{Pe93}, i.e.
 $\sum_{i=1}^M V_i^{\dagger} V_i={\mathbb I}$. Note that the definition
allows the actions on the two subsystems to be correlated; this is how
classical communication enters. Note also that it is the expected 
entanglement that cannot increase---if we end up with a statistical 
ensemble of states there may be a non-zero probability of increased 
entanglement. 

Sometimes one imposes further requirements

(E4) Additivity: $E(\rho_1 \otimes \rho_2)=E(\rho_1)+E(\rho_2)$,

(E5) Continuity: $E$ is a continuous function of $\rho$,

or

(E5') Asymptotic continuity: $E$ is a continuous function of the
fidelity for $n$ copies of the same
pure state, $|\psi\rangle^{\otimes}$
in the asymptotic limit $n\to \infty$,

\noindent the necessity of which is still disputed in the
literature \cite{Vi98,Ho01,DHR01}. Condition (E4) is most welcome  but
difficult to prove in the general case of arbitrary density matrices,
while condition (E5) is satisfied by many different measures. On the
other hand, condition (E5') is rather strong:
assuming  that the measure of entanglement fulfills some technical
conditions which quantify asymptotic continuity one may show
\cite{Vi98,Ho01}
that for pure states it is proportional
   to the entropy of entanglement, $E(|\psi\rangle \langle \psi|)
  = H_N(|\psi\rangle)=-{\rm tr}(\rho^A)\ln (\rho^A)$.
Another set of axioms for the measure of entanglement which singles
out the entropy of entanglement is given by Rudolph \cite{Ru01}.
Such requirements are indeed appropriate in the asymptotic regime but 
it remains interesting to investigate the regime of one copy 
where only the first three conditions are imposed. 
One of the aims of this paper is to emphasize 
that there exist several reasonable measures of entanglement, which
for pure states are different from the entropy of entanglement.

\medskip

\subsection{Quantum R{\'e}nyi entropies}

Consider an $N$ dimensional vector $\vec x$ with non--negative
components normalized as $\sum_{i=1}^N x_1=1$.
The distribution of $x_i$ may be described by
the Shannon (information) entropy
 $H_N:= -  \sum_{i=1}^N x_i \ln x_i$.
A more general family of quantities characterizing the
components of $\vec x$ is provided by
the R{\'e}nyi entropies \cite{Re61} often used in 
information theory \cite{Ka94};
\begin{equation}
H_{\alpha}(\vec x) :=
\frac {1}{1-\alpha}
\ln
 \Bigl[ \sum_{i=1}^N x_i^{\alpha}\Bigr].
\label{Salpha}
\end{equation}
As usual,  in the definition of entropies we adopt the convention
that $0 \ln 0 =0$, if necessary.
The non--negative number $\alpha \ne 1$ is a free parameter labeling
the generalized entropy.
It is easy to see that $\lim_{\alpha\to 1} H_{\alpha}=H_N$,
so for consistency we will write sometimes $H_1$ for $H_N$.
The quantities $H_{\alpha}(\vec x)$
vary from zero (for $\vec x$ with one non--zero component)
to $\ln N$, for the vector ${\vec x}_*$ with all components equal,
$x_i=1/N$. It is possible to show that the generalized entropy is a
non--increasing function of its parameter: for any $\vec x$ and
$\alpha_2 > \alpha_1$,   the inequality
$H_{\alpha_2}({\vec x}) \le H_{\alpha_1}({\vec x})$ holds \cite{Ka94}.

 Some special cases of $H_{\alpha}$  are of particular
interest. Let $r$ denote the number of non-zero components of the
vector $\vec x$,  $x_1$ be its largest component, and
$|{\vec x}|=(\sum_{i=1}^r x_i^2)^{1/2}$ be its length.
Then
\begin{equation}
H_{0}({\vec x})=  \ln r, \quad \quad
H_{2}({\vec x}) =  -\ln |{\vec x}|^2, \quad \quad
H_{\infty}({\vec x}) =  -\ln x_1.
\label{S02inf}
\end{equation}

\begin{figure} [htbp]
   \begin{center}
\
 \vskip -0.5cm
 \includegraphics[width=9.5cm,angle=0]{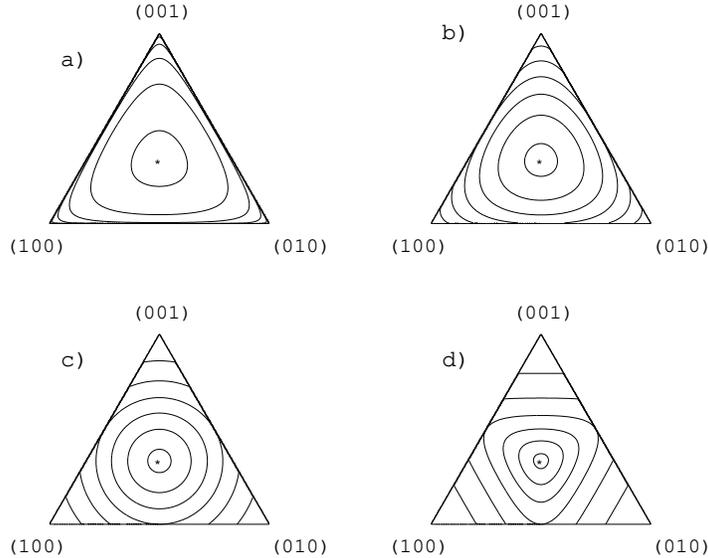}
\vskip 0.3cm
\caption{
 Iso-entropy curves $H_{\alpha}({\vec x})$ in the space of
 $N=3$ vectors $\vec x$
 plotted for
(a) $\alpha=1/3$,
(b) $\alpha=1$ (also contour lines of the Shannon entropy),
(c) $\alpha=2$ (circles of points equidistant from
 ${\vec x}_*$), and  (d) $\alpha=5$.
}
 \label{fig1}
\end{center}
 \end{figure}

To get some experience with the properties of generalized entropies let
us look at Fig. \ref{fig1}, which shows
the iso-entropy curves for $N=3$. The equilateral
triangle, of side length $\sqrt{2}$,
lies in the plane $x_3=1-x_2-x_1$ and its corners are labeled
by  the co-ordinates $(x_1,x_2, x_3)$.
At these three points the generalized entropies
attain their minima, $H_{\alpha}=0$,
for any values of $\alpha\ge 0$.
 The maximum  $H_{\alpha}=\ln 3$ is achieved at the center
of the triangle, which represents the uniform vector
${\vec x}_*$.
The maximum is rather flat for $\alpha=1/3$, as shown in Fig.
\ref{fig1}a. For even smaller, non-negative values of $\alpha$
the isoentropy lines become parallel to the closest side of the
triangle. The entropy $H_0$  vanish at
the corners of the triangle, is equal to
$\ln 2$ at its sides and equals to $\ln 3$ for any point
inside the triangle.
The other example, $\alpha=5$, presented in Fig.
 \ref{fig1}d.
is similar to the limiting case $H_{\infty}$,
for which the iso-entropy curves are parallel to the most
distant side of the triangle.

For any two values of $\alpha$ the isoentropy lines intersect. Thus
any pair of two different generalized entropies induces a different
ordering of vectors $\vec x$.
In our considerations the vector $\vec x$ will represent two
entirely different objects:

\noindent
a)  the spectrum $\vec d$ of a $N=3$ density matrix $\rho$

 or

\noindent
b) the vector of the Schmidt
coefficients $\vec \Lambda$ of a pure state of the $3 \times 3$
composite system.

\medskip

 Thus the pictures of isoentropy lines may
have two entirely different meanings.
In the former case a), the entropy  $H_N$ is equivalent to the
 {\sl von Neumann entropy},
$S_N(\rho) = - {\rm Tr}[ \rho \ln \rho]$, while other
quantum Renyi entropies
\begin{equation}
S_{\alpha}(\rho) = \frac{1}{1-\alpha} \ln {\rm tr} \rho^{\alpha}
\label{quaren}
\end{equation}
 play the role of different measures of
degree of mixing \cite{We78,HHH96}.
The entropy $S_0$ is a function of the rank $r$ of
the density matrix, i.e. the number of positive eigenvalues of
$\rho$. Thus it is not a continuous function of its argument
 in contrast to generalized entropies
$S_{\alpha}$ with $\alpha>0$.
 For $\alpha=2$ we have $S_2 =-\ln [ {\rm Tr} \rho^2]=\ln R$,
where the {\sl purity} (also called {\sl inverse participation ratio})
$R(\rho) = [ {\rm Tr} \rho^2] ^{-1}$
 describes the "effective number of states"
involved in the mixture $\rho$ and varies from unity (for pure
states) to $N$ (for $\rho_*={\mathbb I}/3$).
The purity of any state depends only on its Hilbert-Schmidt
 distance
from the maximally mixed state $\rho_*$,
represented by the center of the triangle.
Thus the curves of constant $R$ have the shape of a circle,
and the corners -- the points  most distant form $\rho_*$ --
denote three orthogonal pure states.
Note that due to the presence of the logarithm in the definition
(\ref{quaren}) the generalized entropies are
additive for product states,
\begin{equation}
S_{\alpha}(\rho_1 \otimes \rho_2)= S_{\alpha}(\rho_1)+ S_{\alpha}(\rho_2)
 \label{additiv}
\end{equation}
for any $\alpha \ge 0$.

In the latter case
b) the centers of the triangles plotted in Fig. 1
represent the maximally entangled states
$|\psi_*\rangle=(|11\rangle+|22\rangle+|33\rangle)/\sqrt{3}$ and the
corners  denote separable states.
The generalized entropy of a vector of the Schmidt coefficients
equals the Renyi entropy of the reduced density matrices,
$H_{\alpha}(|\psi\rangle)=
S_{\alpha}[{\rm tr}_B(|\psi\rangle \langle \psi |)] =
 S_{\alpha}[{\rm tr}_A(|\psi\rangle \langle \psi |)]$.
In particular
 $H_N$ coincides with the entropy of entanglement of the corresponding
pure states of the composite $3 \times 3$ system. As discussed later,
all generalized entropies $H_{\alpha}$ for $\alpha\ge 0$ fulfill the
conditions (E1)-(E3),
and may thus
serve as legitimate measures of the pure state entanglement.
Moreover, for any product pure states these measures are additive,
$ H_{\alpha}(|\psi_1\rangle \otimes |\psi_2\rangle) =
 H_{\alpha}(|\psi_1\rangle)+H_{\alpha}(|\psi_2\rangle)$.
This property, following from (\ref{additiv}),
was pointed out by Vidal \cite{Vi98}.

 Note that an analogous construction
of isoentropy hypersurfaces in the $(N-1)$ dimensional simplex
may represent mixed states acting in ${\cal H}_N$
or pure states of the composite $N \times N$ system.
For any $N\ge 3$ the entropies $H_{\alpha}({\vec x})$ provide not
equivalent measures of uniformity/disorder in the distribution of
the components $x_i$.

\medskip

\subsection{Majorization}

\begin{figure} [htbp]
   \begin{center}
\
 \vskip -1.5cm
 \includegraphics[width=9.0cm,angle=-90]{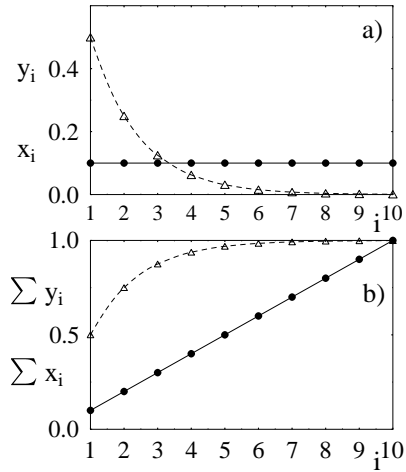}
\vskip -1.5cm
\caption{
 Idea of majorization: (a) the ordered vector ${x}^{\downarrow}=\{x_1,\dots
,x_{10}\}$ ($\bullet$)
is {\sl majorized} by ${\vec y}=\{y_1,\dots ,y_{10}\}$ ($\triangle$),
since the corresponding distribution function takes larger values
(b), and both functions do not cross.
}
 \label{fig:mixi2}
\end{center}
 \end{figure}

The question, which of any two vectors
is  more 'mixed' (or more disordered)
can be approached by the theory of {\sl majorization}
\cite{MO79},
which allows one to introduce a partial order into the
this set.
Consider two  vectors $\vec{x}$ and $\vec{y}$, consisting of $N$
non-negative components each. We
order the components of both vectors in a
decreasing order, (what is sometimes emphasized by the notation 
$ x^{\downarrow}$),
and assume that they are normalized in the sense
$\sum_{i=1}^{N}x_{i}=\sum_{i=1}^{N}y_{i}=1$.
We say that $\vec{x}$ {\sl is majorized} by $\vec{y}$,
 written ${\vec{x}} \prec {\vec{y}}$, if
\begin{equation}
\sum_{i=1}^{k}x_{i}\leq \sum_{i=1}^{k}y_{i}  \label{major}
\end{equation}
for $k=1,2,\dots ,N-1$.
Since the sum of all components of both vectors are equal,
we obtain an equivalent formulation of majorization
\begin{equation}
\sum_{i=l}^{N}x_{i}\geq \sum_{i=l}^{N}y_{i}
\label{Pmajor2}
\end{equation}
for all $l=2,\dots ,N$.
Vaguely speaking, the vector $\vec{x}$ is more 'mixed'
than the vector, $\vec{y}$ and its distribution function grows slower
with the index $i$ (see Fig. \ref{fig:mixi2}).

\vskip -0.5cm

The functions which preserve the majorization
 order are called {\sl Schur convex},
\begin{equation}
{\vec x} \prec {\vec y} \quad {\rm implies } \quad f(\vec x) \le f(\vec
y).
\label{shur}
\end{equation}
Examples of Schur convex functions include
 $f_q({\vec x})=\sum_{i=1}^N x_i^q$ for any $q\ge1$,
 ${\tilde f}_q({\vec x})=-\sum_{i=1}^N x_i^q$ for any
$0< q<1$,
 and $f_e({\vec x})=\sum_{i=1}^N x_i \ln x_i$
\cite{MO79},
while these functions with reversed sign are {\sl Schur concave}.
To show a simple application
of the theory of majorization
let us then consider two density matrices
$\rho_x$ and $\rho_y$, with spectra
${\vec x}$ and ${\vec y}$, respectively.
If ${\vec x} \prec {\vec y}$
then due to Schur-convexity
$H_{\alpha}({\vec x}) \ge H_{\alpha}({\vec y})$
for $\alpha\ge 0$,  since at $\alpha=1$  the prefactor
$1/(1-\alpha)$ in (\ref{Salpha}) changes the
sign and the direction of the inequality is reversed.

A classical theorem by Hardy, Littlewood and Polya
states that ${\vec x} \prec {\vec y}$ if and only if
there exists a bistochastic matrix $B$ such that
${\vec x}=B{\vec y}$. A matrix $B_{ij}$ with non-negative elements is
called {\sl bistochastic} if sum of all elements in each row and
each column equals to unity.
As shown later by Horn \cite{Ho54}
this theorem can be strengthened by replacing the word
bistochastic by unistochastic. A bistochastic matrix
$B$ is called
{\sl unistochastic}, if there exists an unitary matrix $U$
such that $B_{ij}=|U_{ij}|^2$.
All $N=2$ bistochastic matrices are unistochastic, but this is not
true for $N\ge 3$ \cite{MO79}.

Another useful feature of majorization is
 related to the so called {\sl $T$--transform}.
It is a matrix of size $N$ which acts as identity on all but two
dimensions where it has the form
of a bistochastic matrix
\begin{equation}
{\tilde T} =  \left[
\begin{array}{cc}
t  & 1-t  \\
 1-t & t
\end{array}
\right] .
\label{ttransform}
\end{equation}
so $t\in [0,1]$.
One can prove \cite{MO79,Bha97} that
 ${\vec x} \prec {\vec y}$ if and only if
there exist a set of $T$--transforms $T_1, T_2,...,T_k$
that $x=T_1T_2\cdots T_k y$. The maximal number of the
$T$--transforms
needed is not larger than $N-1$. A simple proof of this fact and
of the Horn lemma is recently provided by Nielsen \cite{Ni99b}.

Sometimes it is useful to compare non normalized vectors,
such that
 $\sum_{i=1}^{N}x_{i}\ne \sum_{i=1}^{N}y_{i}=1$.
In such a case one introduces the relation
of {\sl weak majorization} \cite{MO79}
defining  ${\vec{x}} \prec_w {\vec{y}}$ if
\begin{equation}
\sum_{i=1}^{k}x_{i}\leq \sum_{i=1}^{k}y_{i}
\quad {\rm for} \quad k=1,2,\dots,N.
 \label{weakmajor}
\end{equation}
Then vector $\vec x$ is said to be {\sl weakly submajorized} by $\vec y$,
or alternatively, weakly majorized from below.

Majorization technique allows one to obtain effective necessary
criteria for
separability of mixed states, as recently shown by Nielsen and Kempe
\cite{NK00}: If a mixed state $\rho$ is separable then
$\vec{d} \prec {\vec d}_A$ and $\vec{d} \prec {\vec d}_B$
where ${\vec d}_A$ and  ${\vec d}_B$ denote the spectra of the reduced
operators $\rho^A$ and $\rho^B$.
 Since the R{\'e}nyi entropies are Schur concave,
one obtains
\begin{equation}
 S_{\alpha}(\rho^A) \le S_{\alpha} (\rho)
~~~~ {\rm and} ~~~
S_{\alpha} (\rho^B) \le S_{\alpha} (\rho)
\label{entinequal}
\end{equation}
for any separable $\rho$ and $\alpha\ge 0$, which generalizes
earlier results of Horodeccy \cite{HHH96}.

\medskip

\section {Pure states entanglement}
\subsection{Distance to the closest separable state}

For any entangled state $\rho_{ent}$ the distance to the
closest separable state $\rho_c$ may be considered
as a measure of entanglement.
 Each metric in the space of mixed quantum
states will thus define a certain measure of entanglement.
Vedral and Plenio have shown \cite{VP98}
that the Bures distance \cite{Bu69},
$D_B(\rho_1,\rho_2)^2=2-2{\rm tr}|\rho_1^{1/2}\rho_2\rho_1^{1/2}|$,
 to the closest separable state fulfills the
required conditions $(E1)-(E3)$.
For a pure state written in the Schmidt decomposition as
in ( \ref{VSchmidt})
the closest  separable state is
\begin{equation}
\rho_c=\sum_{k=1}^N\lambda_k|kk\rangle\langle kk| \ ,
\label{rhoc}
\end{equation}
(this result is proved in \cite{VP98} for $N=2$ and stated for
larger $N$).
The state $\rho_c$ is typically mixed
 and its squared Bures distance to the pure state
$\rho_{\psi}=|\psi\rangle \langle\psi|$ reads
\begin{equation}
D^2_B(\rho_{\psi},\rho_c)
=2-2\sqrt{ \sum_{k=1}^N \lambda_k^2}
=2-2\sqrt{1/R}  .
\label{Bur2}
 \end{equation}
The inverse participation ratio
$R$ is closely related with the generalized entropy of order $2$,
namely $R=\exp(H_2)$. Therefore the circles around the maximally
entangled state, forming the isoentropy lines for $H_2$ (see Fig. 1c),
form also the lines of the same Bures entanglement. The same concerns 
{\sl generalized concurrence}, another measure of entanglement recently 
introduced by Rungta et al. \cite{Ru01}.

Vedral and Plenio discussed also the {\sl quantum relative
entropy}, $S(\rho_1|\rho_2):={\rm tr}[\rho_1(\ln \rho_1-\ln\rho_2)]$.
Although this quantity does not induce a true metric (e.g. it is not
symmetric), the entropy $S(\rho_{ent}|\rho_c)$ satisfies
$(E1)-(E3)$ and may be considered as a measure of entanglement.
Moreover, they showed that the state (\ref{rhoc}) is also the 'closest'
separable state to the pure state $\rho_{\psi}$
and the smallest relative entropy
$S_{\rm min}=\sum_{k=1}^N \lambda_k \ln \lambda_k$
coincides with the Shannon entropy $H_N$ of the vector of the Schmidt
coefficients.
This quantity, often briefly called {\sl entanglement} of
the pure state $\rho_{\psi}$,
is given a special physical meaning, since the probability of not
distinguishing between
the analyzed entangled state $\rho_{ent}$ and the closest separable
state $\rho_c$ after $n$ measurements behaves as
$\exp[-nS(\rho_{ent}|\rho_c)]$ \cite{VP98}.
Furthermore, the {\sl entanglement of formation}
\cite{BDSW96,Wo98}  of any mixed state
$\rho$, defined as the minimal average entanglement of pure states,
the mixture of which generates $\rho$,
for a pure state reduces
to the entropy of entanglement $H_N$.

For any entangled pure state (\ref{VSchmidt})
one may also look for the closest separable pure state. This problem was
recently considered by Lockhart and Steiner \cite{LS00}, who
show that the state
$|\phi\rangle:= |1'1''\rangle \langle 1'1''|$
 corresponding to the largest Schmidt coefficient
$\lambda_1:={\rm max}\{\lambda_i \}$ is the closest.
They were using the Hilbert--Schmidt
distance, $D_{HS}(\rho_1,\rho_2)=[{\rm tr} (\rho_1-\rho_2)^2]^{1/2}$,
and found
$D_{HS}(\rho_{\psi},\rho_{\phi})=[2-2\lambda_1]^{1/2}$.
However, the projective cross-ratio
$\kappa:=|\langle\psi|\phi\rangle|^2=\lambda_1$ is the only
parameter determining the standard distances
in the space of pure state, e.g. the Bures distance
$D_{B}(\rho_{\psi},\rho_{\phi})=
[2( 1 - \sqrt{\kappa})]^{1/2} $,
the trace distance
$D_{tr}(\rho_{\psi},\rho_{\phi})=2[1-\kappa]^{1/2}$,
and the Fubini--Study distance
$D_{FS}(|\psi\rangle,|\phi\rangle)= \frac{1}{2}
\rm{arccos}(2\kappa-1)=\rm{arccos}(\sqrt{\kappa})$.
All these functions are monotone, so
the measure of entanglement defined as the distance
(any of the above) to the closest separable
pure state defines an ordering of the pure states identical with that
given by the Chebyshev--like like entropy $H_{\infty}=-\ln \lambda_1$.
We have thus shown that three different
settings of the problem of finding
for any pure state of an $N \times N$ system
the closest separable state
reduce to the generalized entropies with
$\alpha= 1,2$ and $\infty$.

 On the other hand one should not expect
that {\sl every} reasonable measure of pure states entanglement
satisfying conditions (E1)--(E3) is neccessarily a function of one of
the entropies (\ref{Salpha}). As a counterexample
let us mention the coefficients of the
characteristic polynomial of the  nontrivial block
of the Gram matrix introduced in \cite{KZ00}.
They can be expressed as symmetric functions
of all Schmidt coefficients, e.g.
 $\sum_{i>j}\lambda_i\lambda_j$,
$\sum_{i>j>l}\lambda_i\lambda_j\lambda_l$,...,
where the summation goes over all possible
sets of $k$ indices, $k=2,...,N$ \cite{Si01}.
All these functions are Schur--concave \cite{MO79},
and thus entanglement monotones, although
in the general case, (for $N\ge 3$)
they are not functions of the generalized entropies.

\medskip

\subsection{$2 \times 2$ system - only one ordering of pure states }

For pedagogical reasons we start to analyze the pure state entanglement
with the simplest $2 \times 2$ system.
Pure states of this $N=4$ system may be parametrized as
\begin{equation}
| \psi_4 \rangle =
(\cos \vartheta_3, \sin \vartheta_3 \cos \vartheta_2 e^{i
\varphi_3}, \sin \vartheta_3 \sin \vartheta_2 \cos\vartheta_1 e^{i
\varphi_2}, \sin \vartheta_3 \sin\vartheta_2 \sin \vartheta_1 e^{i
\varphi_1} ),
 \label{param3}
\end{equation}
where $\vartheta_k  \in [0,\pi/2],$ and $\varphi_k \in [0, 2\pi)$
for $k=1,2,3 $.
The states $| \psi \rangle$
 belong to the complex projective manifold
${\mathbb C}P^3 $. This space is compact
and has $6$ real dimensions, e.g. three polar angles $\vartheta_i$
and three azimuthal angles $\varphi_i$.

To get a better understanding of
${\mathbb C}P^3 $, let us recall
the structure of the space of pure states for systems of even lower
dimension. The set of the $N=2$ pure states,
$| \psi_2 \rangle =
(\cos \vartheta, \sin \vartheta e^{i \phi})$,
is described by the {\sl  Bloch sphere} ${\mathbb C}P^1 \sim S^2$.
The sphere may be drawn in a simplified way by an interval
(a meridian labeled by $\vartheta \in [0,\pi]$), each point of which
represents a circle (a parallel $\vartheta=$const with
 $\varphi \in [0,2\pi)$).
At both poles the circles reduce to a point - see Fig.2.

The $4$--dimensional manifold ${\mathbb C}P^2$
 of the $N=3$ pure states can be  parametrized as
$| \psi_3\rangle= (\cos \vartheta_2,
\sin \vartheta_2 \cos \vartheta_1  e^{i \varphi_2},
\sin \vartheta_2 \sin \vartheta_1  e^{i \varphi_1} )$
where $\vartheta_1 \in [0,\pi/2],$
 $\vartheta_2 \in [0,\pi/2)$
and $\varphi_1, \varphi_2 \in [0, 2\pi).$
These local coordinates allow us to describe (almost all of)
this space as shown in Fig.2b.
The angles $(\vartheta_1, \vartheta_2)$,
describe a point in the positive octant of a sphere $S^2$
(or in an equilateral triangle, what is  topologically equivalent)
which represents entire $2$-torus formed of both phases $\varphi_i$
 \cite{AMM97}.
Each point on one the three edges of the octant
represents a circle, so each entire edge corresponds to a sphere.
Note that three corners of the triangle are
not the 'corners' in  ${\mathbb C}P^2$ --
in the same sense as for $N=2$ the poles (the edges
of the meridian) are topologically equivalent to all other points on the
sphere.

\begin{figure}[htbp]
   \begin{center}
\
 \includegraphics[width=9.5cm,angle=0]{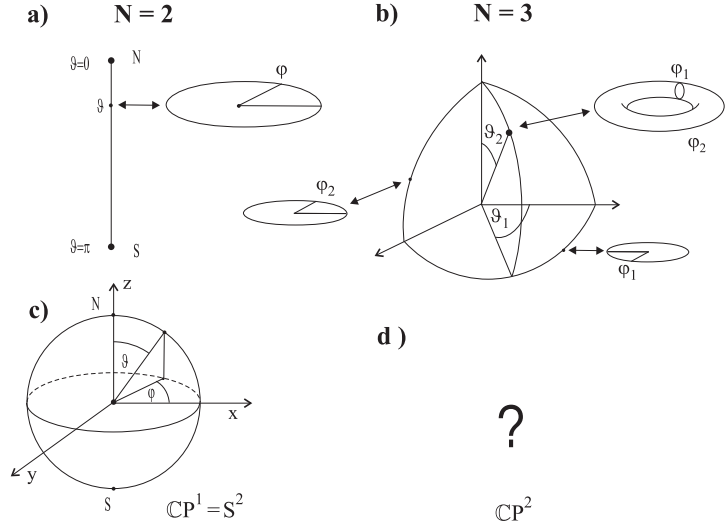}
\vskip 0.3cm
\caption{
The Bloch sphere, $S^2={\mathbb C}P^1$,
of all pure states of a two levels system (a)
may be drawn schematically as a line, each point of which represents a
circle (c). In the same, simplified manner we
depict ${\mathbb C}P^2$ -- the set of all pure states
of the $N=3$ system (b).
Each point inside the octant of a two-sphere,
associated with the angles ($\vartheta_1, \vartheta_2$),
 represents a torus $T^2$ spanned by the
phases ($\varphi_1$, $\varphi_2$).
Each point at the edges of the
octant denotes a circle, so each of three edges represents a sphere.
The questionmark represents the $4$ dimensional manifold
${\mathbb C}P^2$ that we are not able to reproduce
exactly in the picture.
}
   \label{fig3}
\end{center}
     \end{figure}

 These two examples help us to imagine the structure of
 the space ${\mathbb C}P^{3}$ of the $N=4$ pure states.
It may thus be represented as
a $1/16$ part of the hypersphere $S^{3}$,
parametrized by the angles
$\vartheta_k$. Each point inside such a 'hyper-octant'
(or simpler, the  tetrahedron)
represents a $3$--torus spanned by the phases $\varphi_k$,
each point on the face of the simplex a $2$--torus and each point of the
edge a circle.
In general all points of ${\mathbb C}P^{3}$
are equivalent. This symmetry becomes broken if
out of these $N=4$ system we distinguish
two two-levels subsystems.

The corners of the tetrahedron represent
mutually orthogonal states; let us use the standard notation and label
them $(--),(-+),(+-)$, and $(++)$. In principle one could find
analytically
the entropy of entanglement $H_1$ (or any other generalized entropy
$H_{\alpha}$) for each point of the simplex (and each choice of the
phases $\varphi_k$). To get a more transparent picture we prefer
to plot $H_1(|\psi\rangle)$ at the {\sl surface} of the tetrahedron -
see Fig.4.

Four corners represent the separable states.
The same concerns  four edges
of the simplex irrespective of the phases running along the circles.
 Find two maximally
entangled Bell states proportional to $|+-\rangle+|-+\rangle$
and $|++\rangle + |--\rangle$ localized in the middle of two
non-connected edges. The other two orthogonal entangled states,
 $(|+-\rangle+|-+\rangle)/\sqrt{2}$
and $(|++\rangle - |--\rangle)/\sqrt{2}$,
contain non-zero phase, (minus sign) and do not belong to this plot.
The structure of the picture changes with other choices of the basis.
If one uses the Bell basis of
 the four maximally entangled states
and places them into four corners then
the separability of the edge states
depends on the phases $\varphi$.

One may portray any other generalized entropy
$H_{\alpha}(|\psi\rangle)$ in a similar plot. However, in this case of
pure states of the $2 \times 2$ system, all
measures of the entanglement are equivalent in the sense that
one measure is a function of the other one. Therefore
any two entanglement measures, $E_1$ and $E_2$,
generate the same ordering \cite{EP99}
\begin{equation}
E_1(\rho_1) > E_1(\rho_2)
\Leftrightarrow
E_2(\rho_1) > E_2(\rho_2)
 \label{order}
\end{equation}
for any density operators representing pure states,
$\rho_i=|\psi_i\rangle\langle \psi_i|$.
This is due to the fact that for $N=2$ the entanglement of any pure
state is completely characterized by only one relevant parameter --
the  Schmidt angle
$\beta\in[0,\pi/4]$, such that $\lambda_1=\cos^2(\beta)$ while
$\lambda_2=\sin^2(\beta)$.
The distribution of the Schmidt angle for random pure states
distributed according to the natural, unitarily invariant measure
on ${\mathbb C}P^3$ is given by $P(\beta)= 3\cos (2\beta) \sin( 4 \beta)$.
A simple integration allows us to find 
the mean angle, $\langle \beta \rangle_{{\mathbb C}P^3}=1/3$
which incidentally equals the mean entropy of entanglement, 
$\langle H_1(|\psi\rangle)\rangle_{{\mathbb C}P^3}=1/3$
\cite{Ha98,ZS00}.

\vskip -1.0cm
\begin{figure}[htbp]
   \begin{center}
\
 \vskip 0.3cm
 \includegraphics[width=8.5cm,angle=0]{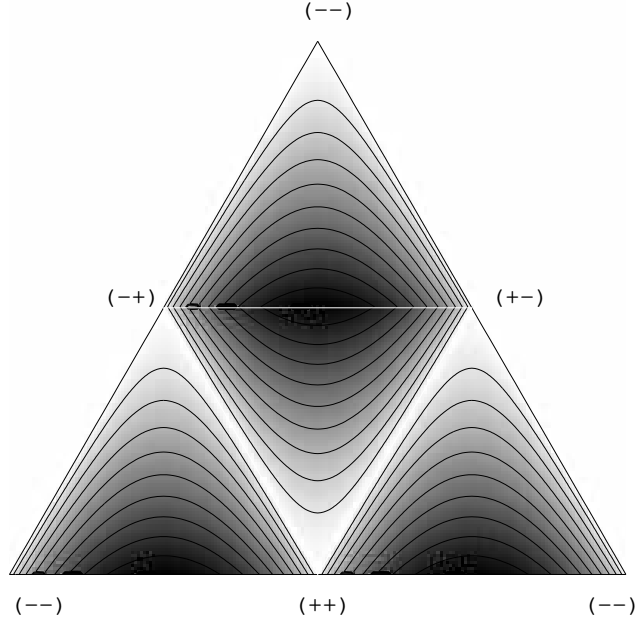}
\vskip 0.3cm
\caption{Entropy of the entanglement, $H_1(|\psi\rangle)$
for pure states of the $2 \times 2$ problem
at the {\sl surface} of the $\vartheta$--tetrahedron representing
${\mathbb C}P^3$
(dark colors denote large entanglement).
Labels refer to the
corners representing the standard basis.
Copy this picture magnifying, cut out the triangle, fold three times
along the edges and glue together into a thetrahedron
 to enjoy the symmetry
 of this representation.
}
   \label{fig4}
\end{center}
     \end{figure}

\medskip

\subsection{Global transformations of mixed states}

Before dealing with the more complicated case of entanglement for the $3
\times 3$ system let us make a detour to have a closer look at
the space of mixed quantum states acting in ${\cal H}_N$.
Any state $\rho$ may brought to a diagonal form by a
unitary evolution $\rho\to \rho'=U\rho U^{\dagger}$
which preserves the vector of eigenvalues
$\vec d=\{d_1,d_2,...,d_N\}$. The normalization
condition assures tr$\rho=\sum_{i=1}^N d_i=1$.

Any physical operation can be described
by a superoperator ${\tilde M}:\rho\to \rho'$
that preserves positivity
of the mixed state $\rho$. Strictly speaking the term
'positive' refers to {\sl positive semidefinite} Hermitian
operators, which do not have negative eigenvalues.
 Moreover, the system may be coupled
to an environment, and $\tilde M$ may be trivially extended to
${\tilde M} \otimes {\mathbb I}$. A superoperator ${\tilde M}$
is {\sl completely positive} if for all such extensions
${\tilde M} \otimes {\mathbb I}$ is positive.
A completely positive (CP)-map
which preserves the trace (normalization of the state is conserved)
is called a {\sl stochastic map} or {\sl quantum channel}.
It is capable to describe
any quantum operation including the process
of quantum measurement and
 may be represented in the {\sl Krauss form}  \cite{Kr71}
\begin{equation}
\rho\to \rho'={\tilde M}(\rho)=
\sum_{i=1}^k A_i\rho A_i^{\dagger}
,
 \label{Krauss}
\end{equation}
where  the set of $k$ operators $A_i$
satisfies the completeness relation,
$\sum_{i=1}^k A_i^{\dagger} A_i={\mathbb I}$.

In the following we shall consider a smaller subset of
quantum channels, which may be written in the form
$\rho'= M(\rho)=\sum_{i=1}^k p_i U_i\rho U_i^{\dagger}$,
where each operator $U_i$ is unitary
and the positive coefficients $p_i$ sum to unity.
These operations, called {\sl external random fields} \cite{AL87},
are {\sl unital}, since the maximally mixed state $\rho_*={\mathbb I}/N$
is preserved, $M(\rho_*)=\rho_*$. Maps which
preserve both the trace and the identity are called {\sl bistochastic}.

Let $\vec{d}'$ denote the vector of eigenvalues of $\rho'=M(\rho)$.
Then one can prove  \cite{Ul71,Ni00}
 the following majorization relation,
  $\vec{d}' \prec  \vec{d}$.
Using the concept of the $T$--transforms
we are going to show the converse:
given a vector of eigenvalues ${\vec d}'$ such that
 $\vec{d}' \prec \vec{d}$
there exists a bistochastic map $M:\rho'=M(\rho)$, written
$\rho 
\stackrel{\rm global}{\longrightarrow}  \rho'$.
Thus both properties  are equivalent,

\begin{equation}
\rho 
\stackrel{\rm global}{\longrightarrow}
\rho'
\quad \Leftrightarrow \quad
\vec{d}' \prec  \vec{d}.
 \label{equimix}
\end{equation}

For simplicity  we start discussing
the $N=3$ case, represented in Fig. 5.
Consider a state $\rho$ with spectrum $\vec d$,
(its components are ordered as $d_1\ge d_2\ge d_3$)
and a transformation
$M_{12}(\rho)=w_{12} U_{12} \rho U_{12}^{\dagger}+(1-w_{12})\rho$.
The unitary matrix $U_{12}$ acts as an identity
on all but two first components, (labeling the matrix)
 which get mixed by the
orthogonal submatrix:
$(U_{12})_{11}=(U_{12})_{12}=(U_{12})_{22}=1/\sqrt{2}$ and
$(U_{12})_{21}=-1/\sqrt{2}$ .
In this way $M_1$ moves $\rho$ along the horizontal line
joining $\vec d$ with ${\vec d}_a=( d_{12},d_{12}, d_3)$,
where $d_{12}=(d_1+d_2)/2$.
The length of this move is controlled by the weight parameter $w_{12}$.
An analogous transformation $M_{23}(\rho)$ defined by
the unitary matrix $U_{23}$ which preserves the
first eigenvalue $d_1$, allows one to obtain any state
along the line between $\vec d$  and
${\vec d}_b=(d_1,d_{23},d_{23})$,
where $d_{23}=(d_2+d_3)/2$.
It is easy to see that any state with the spectrum fulfilling
$\vec{d}' \prec \vec{d}$ (shaded region in Fig. 5a),
 may be reached by a superposition
of operations $M_{ij}$, which correspond to $T$--transforms.
Observe that the lines limiting the accessible region are
parallel to the
isoentropy curves $H_{0}$ and $H_{\infty}$, respectively.

\begin{figure} [htbp]
   \begin{center}
\
 \includegraphics[width=17.0cm,angle=0]{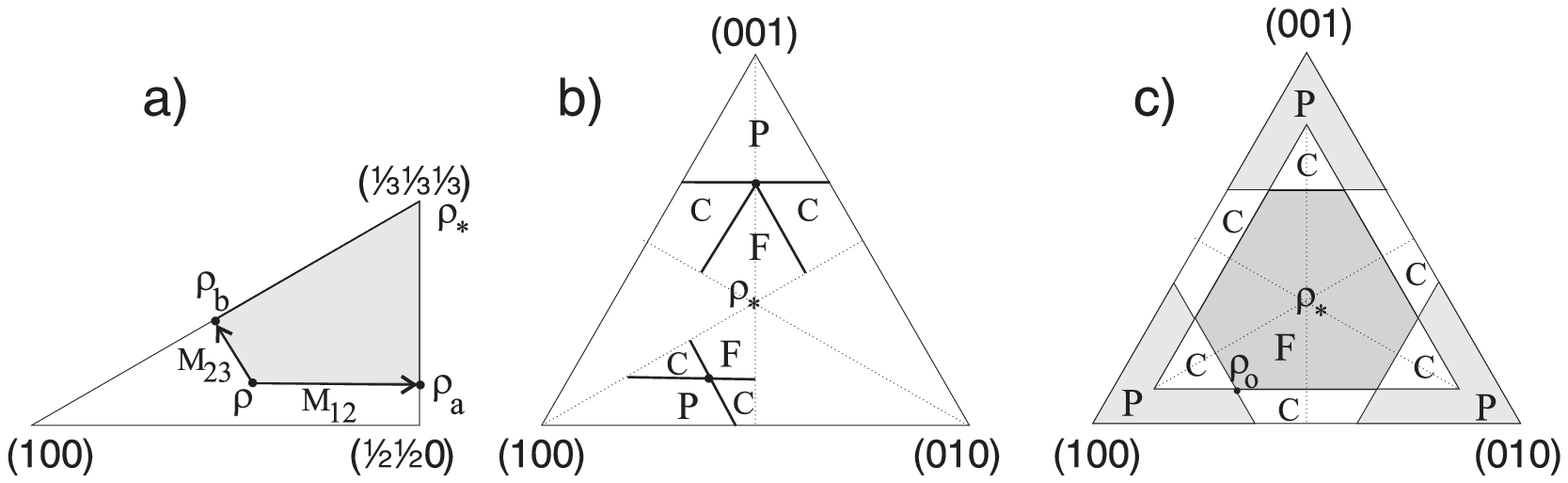}
\vskip 0.6cm
\caption{
 Simplex of eigenvalues for $N=3$ density matrices:
 corners represent pure states,  center the maximally mixed state
 $\rho_*={\mathbb I}/3$;
  (a) asymmetric part of the simplex (Weyl chamber), for which
       $d_1\ge d_2 \ge d_3$. Evolution governed by two
$T$--transforms defines the shaded region accessible from $\rho$;
  (b) the shape of the 'light cone' depends on the
       degeneracy of the spectrum:
        $F$ - denotes {\sl Future}, $P$ {\sl Past},
       and $C$ the {\sl non comparable} states;
   (c) splitting of the entire simplex into three parts
       with permutation operation allowed.
}
 \label{fig5}
\end{center}
 \end{figure}

This construction is simply generalized for
arbitrary $N$: the accessible region of states with
spectra majorized by $\vec d$
is a convex polytope with $2(N-1)$ faces.
The point $\vec d$ sits in one of its corners, which joins
$N-1$ edges
-- lines generated from $\rho$ by the transformations $M_{i,i+1}$,
where $i=1,...,N-1$.
Therefore every face forming the part of boundary
of the accessible region,
(corresponding to the light cone in special relativity),
may be generated by a sequence
of $(N-2)$ T--transforms acting on $\vec d$.
The first transform, $M_{1,2}$, moves $\rho$ along a line
paralell to the closest face of the simplex, so the entropy
$H_0$ is conserved. All other transforms, $M_{i,i+1}$,
do not influence the largest component, $d_1$, so the
entropy $H_{\infty}$ remains unchanged.

The majorization property  $\vec{d}' \prec \vec{d}$ assures that the
Von Neumann entropy and all other entropies $S_{\alpha}$
for $\alpha \ge 0$
grow (strictly speaking do not decrease) during the time
evolution governed by random external fields $M$.
Note that the factor $1/(1-\alpha)$
in the definition (\ref{Salpha}) is crucial to
assure the Schur convexity of $H_{\alpha}$.
For any state $\rho$ it is easy to specify the
set of all states $\rho'$ accessible from it by
external random fields. For these states, denoted in Fig.~5b
by $F$ as 'Future',
{\sl all} generalized entropies satisfy
   $S_{\alpha}(\rho') \ge S_{\alpha}(\rho)$.
The complementary set of all states $'\rho$
which may be transformed by $M$ into $\rho$ is denoted by $P$ as 'Past'.
Obviously $S_{\alpha}(\rho) \ge S_{\alpha}('\rho)$.
There exists also a set $C$ of noncomparable states,
which are not ordered with respect to $\rho$ by the majorization
relation. This structure is thus analogous to the
{\sl special relativity} picture of the light cone.
The actual shape of the 'cone'
(e.g. the size of the angle 'Future') depends
on the degeneracy of $\vec d$
as shown in  Fig. 5b: in the generic case it is equal to $2\pi/3$,
while at a bisetrix it shrinks to $\pi/3$.
The 'world line' crossing the point $\rho$
and representing a possible
trajectory in the space of density matrices
is located inside the accessible region $F$.

The simplex can be divided into $N!$ asymmetric
simplices  (called {\sl Weyl chambers} in group theory),
which differ by the permutation of the
components of the vector $\vec d$.
In the discussed
case $N=3$ three bisetrices split the simplex into $6$
right triangles, one of which is shown in Fig. 5a.
A permutation of the components of $\vec d$,
which might  be considered as a special unitary case of the
random fields $M$, maps $\vec d$ into a corresponding
symmetric point in one of the other $6$ right triangles.
Therefore,
in the generic case of a non-degenerate spectrum of $\rho$,
the entire decomposition of the $N=3$ equilateral
triangle of eigenvalues looks as shown in Fig 5c.
The set 'Future' has the form of a hexagon, which contains two triples of
equal sides,  and becomes regular for ${\vec d}=(3,2,1)/6$, or in
general for ${\vec d}=(1/3+x,1/3,1/3-x)$, where $x\in (0,1/3]$. If
the spectrum of  $\rho$ is degenerate
(the vector $\vec d$ is located at a bisetrix),
the hexagon reduces to an equilateral  triangle.

It is not difficult to understand how this construction works
in higher dimensions: the set 'Future'
is a convex polytope, defined by
the convex hull of $N!$ points
obtained from ${\vec d}$ by permutations of its components.
This set contains $N!$ identical simplices surrounding $\rho_*$
and is topologically equivalent to a ball.
It consists of $N!$ vertices (one for each permutation),
each joining $N-1$ edges.  Thus in the generic case this polytope
has
$N!(N-1)/2$ edges and $2^N-2$ (hyper)faces of dimension $N-2$.
For $N=4$ it is a polyhedron with $24$ vertices, $36$ edges and
$14$ faces; if the vertex at ${\vec d}$ lies on the line
\begin{equation}
{\vec d}(x)= \frac{x}{6}(3,2,1,0) +
             \frac{1-x}{4}(1,1,1,1);
\qquad 0\le x\le 1
 \label{arch}
\end{equation}
 then we obtain an Archimedean solid known as the truncated octahedron
whose surface is composed of $8$ regular hexagons and $6$ squares. By
definition an Archimedean solid has regular but not equal faces
 \cite{Cromwell}.
If the eigenvalues are degenerate, the polytope degenerates as well.
It is amusing to notice that a number of Platonic and
Archimedean solids appear for special choices of ${\vec d}$.
In particular there is a surface in the simplex at which
the degeneracy pattern is  ${\vec d}=(x_0,x,x,x_3)$.
If the vertex sits at a particular line on this surface
then the polytope is the Archimedean cub-octahedron.
Similarly,
the surfaces determined by  ${\vec d}=(x_0,x_1,x,x)$ or
 ${\vec d}=(x,x,x_2,x_3)$ contain lines giving rise to the truncated
tetrahedron. On the lines
${\vec d}=(x_0,x,x,x)$ and ${\vec d}=(x,x,x,x_3)$
we obtain regular tetrahedra while the line
${\vec d}=(x,x,y,y)$ yields regular octahedra. So much about the
'Future'.
The set 'Past' consists of $N$ disjoint parts
occupying neighbourhoods of each corner, while $C$ consists of
$2^N-2$ parts, each for every face of the set 'Future'. 
The flat geometry of the simplex of eigenvalues
has a concrete  meaning, since the Euclidean distance between any two
points of a Weyl chamber (Fig. 5a) is equal to the Hilbert-Schmidt
distance between the global orbits they generate -- see Appendix A.

Note that the analogy with special relativity does not work in the simplest
case of $N=2$,  for which all measures of the degree of mixing are equivalent.
All of them become functions of one single parameter,
(say, the distance to the center of the Bloch ball),
and generate the same order of the density matrices.
The situation changes for $N=3$, for which the spectrum
of $\rho$ depends on two parameters,
 as recently  discussed by  Jozsa and Schlienz \cite{JS00}.
They obtained
interesting relations  between different entropies $H_{\alpha}$,
which may also be proved using the techniques of majorization.
which may also be proved using the techniques of majorization.
For example, if  we move in the triangle of eigenvalues
along a circle of a constant $R$ and tr$\rho^3$ grows
(the entropy $H_3$ decreases), then the von Neumann
entropy $H_1$ increases. Such relations may be
figured out by superimposing together different
isoentropy curves shown in Fig. 1.

\medskip

\subsection{Local transformations of pure states}

Consider now the set of pure states of an $N \times N$
composite system. We divide it into equivalence classes consisting
of states that can be transformed into each other by local unitary
transformations; the resulting orbit space is precisely a Weyl chamber
in the Schmidt simplex.
Let $\Lambda_{\psi}=\{\lambda_1(\psi), ...,\lambda_N(\psi)\}$
represents the vector of the Schmidt coefficients of a state
$|\psi\rangle$ ordered in the descending order. Define a deterministic 
LOCC transformation as one that takes states to states (as opposed 
to one that transforms a state to a statistical ensemble of states). 
Nielsen has shown that a deterministic LOCC transformation of a
given state $|\psi\rangle$ into $|\phi\rangle$ is possible
if and only if the corresponding vectors of Schmidt coefficients
satisfy the majorization relation  \cite{Ni99}

\begin{equation}
|\psi\rangle
\stackrel{\rm LOCC}{\longrightarrow} |\phi\rangle
\quad \Leftrightarrow \quad
\Lambda_{\psi} \prec \Lambda_{\phi} .
 \label{equipur}
\end{equation}
Note a similarity between this relation concerning the {\sl local}
transformation of pure states and
the condition (\ref{equimix})
concerning the {\sl global} bistochastic transformations
of mixed states.

Any pure state $|\psi\rangle$
allows one to split the Schmidt simplex, representing local orbits of
pure states, into three regions:  'Future' $F$, 'Past' $P$, and
non-comparable,  $C$.
This structure for $N=3$ is shown in Fig. 6. Although
Fig. 5 describes an entirely different physical problem,
both picures are  almost identical. The only difference
 consists in the {\sl arrow of time}:
the 'Past' for the evolution in the space of density matrices
corresponds to the 'Future' for the local entanglement transformations
and vice versa. The triangles plotted here should not be confused with
Fig. 4. In fact the analogous plot for $N=2$
would contain only one line, say joining the states
$(++)$ and $(--)$, along which the entanglement
takes all values in $[0, \ln 2]$.
On the other hand, the triangle shown in Fig. 6 may be given a direct
geometrical interpretation, provided it is deformed into an octant
of the sphere. Formally this may be achieved by the transformation
$\{\lambda_1,...,\lambda_N\} \to \{\lambda_1^2,...,\lambda_N^2 \}.$
Then the length of the arc joining two points of the Weyl chamber (which
covers $1/48$ of the sphere) is proportional to the Fubini-Study
distance between the local orbits they generate - see Appendix A.

In general, all different measures of mixing increase
during the operations considered and all the measures of entanglement
decrease. In particular it follows that all generalized entropies
$H_{\alpha}(\Lambda)$ and their convex functions
belong to the class of {\sl entanglement monotones}
which do not increase during local operations \cite{Vi98}.
For example $H_{\infty}(\Lambda)$ equals
minus maximal Schmidt coefficient, $-\lambda_1$, a simple entanglement
monotone discussed by Vidal \cite{Vi98}.
The entropies $H_{\alpha}$ satisfy thus the conditions
(E1)-(E3) required for a measure of the entanglement.
The maximal fidelity of a pure state,
$F_{\rm max}(\psi)=(\sum_{i=1}^N \sqrt{\lambda_i})^2/N$ \cite{VJN00},
 a convex function of the generalized entropy,
$F_{\rm max}=\exp(H_{1/2})/N$, is also an entanglement monotone due to
the Schur convexity. So is another measure of entanglement called
{\sl robustness}, which is defined as
the minimal amount of separable noise that has to be mixed
with the analyzed state to wash out completely its
quantum correlations \cite{VT99,Vi98}. For pure states the
robustness equals $N F_{\rm max}-1 = \exp(H_{1/2})-1$.

Another quantity used to characterize the entanglement
is called {\sl negativity} \cite{ZHSL,Zy99}. It is
based on the concept of partial transposition (transposition in one of
the two subsystems).
 It is known that for any separable state $\rho$ its partial transpose
$\rho^{T_2}$ is positive \cite{Pe97,HHH}, and the negativity
is expressed by the trace norm, $t:=|\rho^{T_2}|_{\rm tr}-1$,
i.e. by the sum of absolute values of the eigenvalues.
Negativity is an entanglement monotone and satisfies the axioms
(E1)--(E3) \cite{VW01}.
For pure states of  $ 2 \times 2$ system
it is equal to {\sl concurrence}, used
by Wootters to calculate the entanglement of formation \cite{Wo98,Wo01}.
For any $N \times N$ pure state written in the Schmidt form
(\ref{VSchmidt}),
 the partially transposed matrix $\rho^{T_2}$ has a simple block structure:
it consists of  $N$ Schmidt  numbers at the diagonal
(which sum to unity) and $N(N-1)/2$ blocks of size $2$,
one for each pair of different indices $(i,j)$.
Eigenvalues of each block are
 $\pm \sqrt{\lambda_i \lambda_j}$, so the
 negativity  reads
\begin{equation}
t(|\psi\rangle) =2 \sum_{i>j=1}^N \sqrt{\lambda_i \lambda_j }
= \Bigl( \sum_{i=1}^N \sqrt{\lambda_i} \Bigr)^2 -1 ,
\label{negat}
\end{equation}
which varies from $0$ for separable states to $N-1$
for maximally entangled states.
It is easy to see that for pure states the
negativity is equal to robustness, so is also
a function of the quantum R{\'e}nyi entropy entropy of order one half,
$t=\exp(H_{1/2})-1$.
Interestingly, the same behaviour for the pure states
is characteristic to the {\sl cross norm} - another measure of
entanglement introduced by Rudolph \cite{Ru00}.
Although negativity $t$ does not satisfy (E4)
its nonlinear function, $H_{1/2}=\ln (t+1)$,
is additive in the general case of arbitrary mixed states,
as all other R{\'e}nyi entropies.
On the other hand, the negativity suffers a serious drawback:
it is not capable to detect so-called
{\sl bound entangled} states with positive partial transpose,
which are known to exist for quantum
systems with dim${\cal H} \ge 8$ \cite{Ho97,HHH98}.

Let us emphasize that the analogy to special relativity
holds for the description of local
transformations of pure states entanglement.
This analogy was put forward explicitly
in the paper by Hwang et al. \cite{HAHL00}
and used (correctly) to describe a related problem
of the mixed states entanglement.
As we point out in this work,
the same analogy is useful also in investigating
the pure states entanglement.
However, the authors of \cite{HAHL00}
discussed only the simplest $2 \times 2 $ problem,
for which all entanglement measures for pure states
generate the same ordering.  In the general case
this is no longer true.
For example the maximal fidelity $F_{\rm max}$ and the
entropy of entanglement $H_1$ generate different ordering
of pure states for $N>2$, as can be deduced in Fig. 1
for $N=3$. The fact that various entanglement measures
 generate different ordering of the set of pure states
(condition (\ref{order}) is violated),
has been raised in \cite{Vi99,VP00}. Vidal concludes
that it is necessary to use $N-1$
different measures of entanglement
to provide a detailed description of
the set of pure states of $N \times N$ system.
This is a consequence of the fact that
there exists exactly $N-1$ independent
Schmidt coefficients in the decomposition (\ref{VSchmidt}).
A possible set of entanglement monotones is
given by the sum of the $N-k+1$ {\sl smallest} Schmidt coefficients,
$E_k=\sum_{i=k}^N \lambda_i$,
with $k=2,...N$ \cite{Vi98}. Note that the sum of $N-1$
smallest coefficients is related to the already mentioned monotone,
the largest coefficient, $\lambda_1=1-E_{2}$.

\vskip  0.2cm
\begin{figure} [htbp]
   \begin{center}
\
 \includegraphics[width=17.0cm,angle=0]{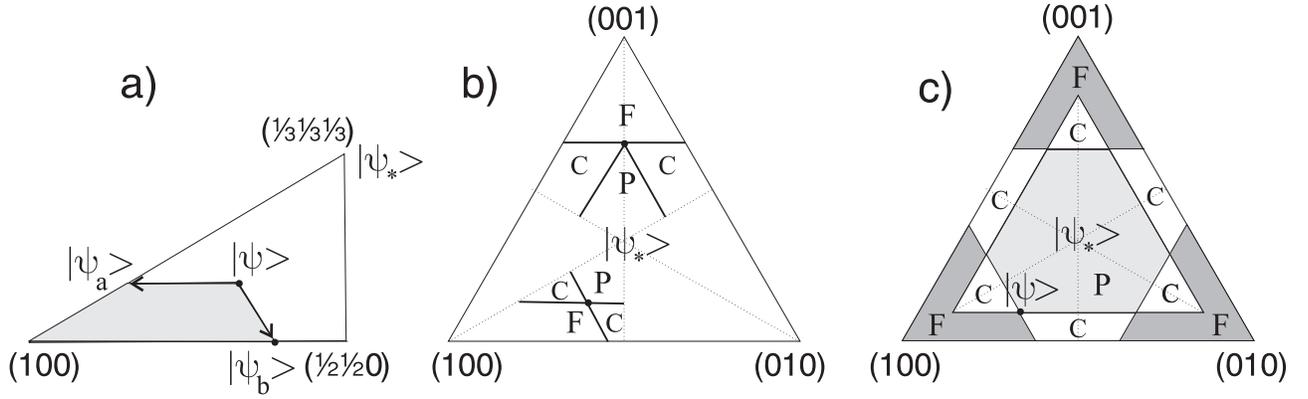}
\vskip 0.5cm
\caption{
 Simplex of Schmidt coefficients for $3\times 3$ pure states:
 corners represent separable states,  center the maximally entangled
state
 $|\psi_*\rangle$. Panels (a-c) analogous to these in Fig. 5, observe
the opposite direction of the arrow of time. As in Fig. 5
the bounding lines are the level curves of $H_0$ and $H_{\infty}$ familiar
from Fig. 1.
}
 \label{fig6}
\end{center}
 \end{figure}

\vskip -0.5cm

Pure states with the same set of the Schmidt
coefficients (e.g. $6$ points in fig. 6c
generated be reflections with respect to all bisetrices)
are called {\sl interconvertible} \cite{JP99}, since they
(and only they) might be transformed reversibly by local transformations.
The generic orbit of pure states equivalent
by local unitary operations has $N^2-N-1$ dimensions,
and this number decreases in the case of a degeneracy
in the vector of the Schmidt coefficients \cite{Si01}.
On the other hand, the 'noncomparable' states
cannot be joined by a local transformation in
any direction. The existence of such pairs of states
(analogous to the space--like events in special relativity)
was recently discussed in \cite{Vi99,Ni99,JP99,VJN00}.
In fact, Nielsen conjectured \cite{Ni99} that the probability
of picking at random
two incomparable states out of the set of all $N\times N$ pure states
(according to the natural, rotationally invariant measure)
tends to unity in the limit $N\to \infty$. We may now support
this conjecture
with a simple geometrical argument: the states close to the faces
of the Schmidt simplex are incomparable, (see Fig. 6c)
and it is well known that for large dimensional simplices
almost all its volume is contained in a thin 'skin'
close to its faces. Furthermore,
the natural, unitarily invariant measure on
${\mathbb C}P^{N-1}$ generates the Hilbert-Schmidt measure
in the Schmidt simplex \cite{Ha98,ZS00},
according to which the degeneracies in $\Lambda$ are avoided:
the probability distribution exhibits maxima in vicinity of the faces of the
simplex, while its minima occur at the center of the
simplex and along the symmetry planes (bisetrices of the triangle)
including the corners.

\medskip

\subsection{Probabilistic transformations of pure states}

Finally we discuss the case when the
majorization condition (\ref{equipur}) is not fulfilled,
so that deterministic local transformation of the pure state 
$|\psi\rangle$ into $|\phi\rangle$ is not possible. It is 
not possible to create entanglement by means of LOCC if the initial state 
is separable, but it can be shown that if the numbers of nonzero components 
in both Schmidt vectors are the same then one may still transform 
$|\psi\rangle$ into $|\phi\rangle$ with  a non-zero probability $P$ of 
success. More precisely it was shown by Vidal
\cite{Vi99} that the optimal protocol yields
\begin{equation}
P(|\psi\rangle \to |\phi\rangle) =
\underset{k\in [1,N]}{\rm min} \ 
\frac{\sum_{i=k}^N \lambda_i(\psi)}
     {\sum_{i=k}^N \lambda_i(\phi)} =
\underset{k\in [1,N]}{\rm min} \ 
\frac{ E_k (\psi)}{ E_k (\phi)} .
\label{prob1}
\end{equation}
This statement may be rephrased
by saing that the optimal probability
$P(|\psi\rangle \to |\phi\rangle)$ is the greatest weight $p$ such that
the submajorization relation holds \cite{VJN00},
\begin{equation}
\Lambda_{\psi} \prec _w p \Lambda_{\phi} .
 \label{prob2}
\end{equation}

\noindent Probabilistic transformations are important because they make 
entanglement purification procedures possible.

Figure 7 shows the probability of accessing different regions of the
Schmidt simplex for pure states of a $3 \times 3 $ system for four
different initial states $|\psi\rangle$. The shape of the black figure
($p=1$ represents deterministic transformations) is identical with the set
'Future' in Fig.6. The more entangled final state $|\phi\rangle$
(closer to the maximally entangled state in the center of the triangle),
the smaller probability $p$ of a successful transformation.
Observe that the contour lines (plotted at $p=0.2,0.4,0.6$ and $0.8$)
are constructed from the iso-entropy lines $H_{\alpha}$ for
$\alpha\to 0$ and $\alpha\to \infty$ (see Fig. 1.)

\vskip  0.2cm
\begin{figure} [htbp]
   \begin{center}
\
 \includegraphics[width=14.0cm,angle=0]{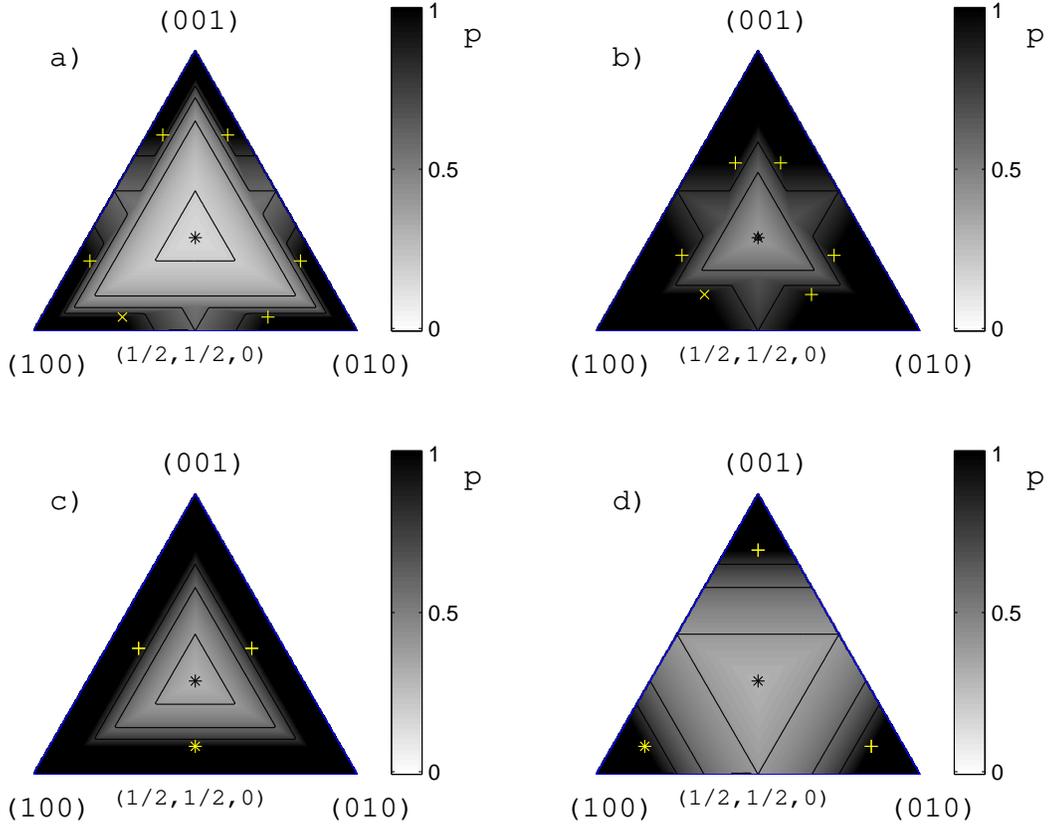}
\vskip 0.5cm
\caption{ Probabilistic transformation of pure states in the
 simplex of Schmidt coefficients for $3\times 3$ pure states:
 white $(\times)$ represents the initial state $|\psi\rangle$\, $(+)$ the
interconvertible states, and black $(*)$ the maximally entangled state.
The gray scale represents the probability $p$, with which
a given state may be obtained by local transformations from $|\psi\rangle$.
Initial state   $|\psi\rangle$\
is characterized by the Schmidt coefficients
$\Lambda=(\lambda_1,\lambda_2,1-\lambda_1-\lambda_2)$, where
$( \lambda_1,\lambda_2)=$
 (a) $(0.7, 0.25)$, (b) $( 0.6, 0.27)$,
 (c) $(0.45, 0.45)$, and (d) $(0.8, 0.1)$.
Due to the degeneracy in $ \Lambda$ in the last two cases there exist only
three different interconverible states.
}
 \label{fig7}
\end{center}
 \end{figure}

\section{Concluding remarks}

We presented a brief review of various measures of entanglement and ways
to define them. Analyzing geometrical properties
of the pure states entanglement we have shown
that there exist several non-equivalent measures
of entanglement which generate different ordering of states.
Four different measures often discussed in the literature:
robustness of entanglement (negativity or cross norm), entropy of
entanglement, Bures distance
to the closest mixed state and the  distance to the closest pure state
are functions of the generalized R{\'e}nyi--like entropies
of the vector of the Schmidt coefficients (\ref{Salpha})
with $\alpha=1/2,1,2$ and $\alpha\to \infty$, respectively, - see Table
1.

\medskip

\hskip 2.4cm
\begin{tabular}
[c]{|l|l|c|}\hline\hline
\multicolumn{1}{||l|}{Measure of} & \multicolumn{1}{||l|}{Formula for} &
\multicolumn{1}{||l||}{Monotone function}  \\
\multicolumn{1}{||l|}{entanglement} & \multicolumn{1}{||l|}{pure states} &
\multicolumn{1}{||l||}{of $H_{\alpha}$ with} \\ \hline\hline
Schmidt rank & number of nonzero $\lambda_{i}$ & $\alpha=0$\\\hline\hline
Maximal fidelity & $F_{\max}=\exp(H_{1/2})/N$ & 
\\\cline{1-2}%
Negativity & $t=\exp(H_{1/2})-1$ & $\alpha=\frac{1}{2}%
$\\\cline{1-2}%
Robustness & $NF_{\max}-1=\exp(H_{1/2})-1=t$ & \\ \cline{1-2}
Cross--norm &  & \\\hline\hline 
Entropy of entanglement & $E=H_N=-\sum_{i=1}^{N}\lambda_{i}\ln\lambda_{i}$ &
\\\cline{1-2}%
Relative entropy  & $\min_{SEP} S(\rho||\rho_s)=E$ & \\\cline{1-2}%
Entropy of formation & $E_{F}=E$ & $\alpha=1$\\\cline{1-2}%
Entropy of distillation & $E_{D}=E$ & \\\hline\hline
Minimal Bures distance &  & \\
to a separable ({\bf mixed}) state & $\min_{SEP}(D_{B})=2[1-\exp(-\frac{1}{2}%
H_{2})]$ & $\alpha=2$\\ \cline{1-2}
Generalized concurrence  &  & \\ \hline\hline
Minimal distance to &  & \\
a separable ({\bf pure}) state & & \\
\cline{1-2}
\ \ \ Fubini-Study distance &
$\min_{SEP}(D_{FS})=\arccos[{\sqrt{\lambda_{\max}}}]$ & \\\cline{1-2}%
\ \ \ \ \ \ \ \ \ \ trace distance & $\min_{SEP}(D_{tr})=2\sqrt{1-\lambda
_{\max}}$ & $\alpha\rightarrow\infty$\\\cline{1-2}%
\ \ \ \ Hilbert-Schmidt distance & $\min_{SEP}(D_{HS})=\sqrt{2(1-\lambda
_{\max})}$ & \\\cline{1-2}%
\ \ \ \ \ \ \ \ Bures \ \ distance & $\min_{SEP}(D_{B})=\sqrt{2(1-\sqrt
{\lambda_{\max}})}$ & \\\hline\hline
\end{tabular}

\smallskip
{\bf Table. 1.} Measures of the entanglement for $N \times N$
bi--partite system ordered according to the
properties at the manifold of pure states.

\bigskip

Out of these measures only the entropy of entanglement 
is additive, if pure states are concerned. The remaining
quantities (not equivalent to $S$) are not, but each measure may be related
with a certain additive generalized entropy $H_{\alpha}$.
Thus no single measure seems to be a priori distinguished. On the other hand, 
some of them may be preferred by certain experimentally feasible
protocols of local manipulations of the pure states,
(e.g. the distillation of entanglement), or by the
property (E5') of the asymptotic continuity  \cite{Vi98,DHR01}.

The above analysis has direct consequences
for the more complicated problem of the entanglement of mixed states.
Consider the definition of {\sl entanglement of formation}
of a mixed state \cite{BDSW96}
\begin{equation}
EoF(\rho):= {\rm min} \sum_{i=1}^L p_i E(|\psi_i\rangle
), ~~~ ~~~ p_i> 0 ,
\end{equation}
where the minimum is taken with over all
pure states decompositions
$\rho=\sum_{i=1}^L p_i |\psi\rangle \langle \psi|$
of a finite length $L$,
normalized by $\sum_{i=1}^L p_i=1$.
In this expression one makes use of the entropy of
entanglement,
$E(|\psi\rangle)=H_N(|\psi\rangle)$,
but one may also apply
other generalized entropies $H_{\alpha}(|\psi\rangle)$
to generate reasonable
measures of the mixed states entanglement \cite{Vi98}.
It is easy to see that these generalized
entropies of formation induce
different orderings in the set of density matrices.

Time evolution of pure states under
such transformations reveals an
analogy to special relativity: We first divide the set of pure states
into equivalence classes consisting of the orbits under local unitary
transformations. Then we find that for each point in this space of
equivalence classes of pure states there exists a
'light cone' which divides the space into the 'Future' (F), the 'Past'
(P) and the non--comparable states (C).
In the Schmidt simplex the role of
the light cone is played by the accessible region limited by
(hyper)faces spanned by any set of $(N-2)$ edges determined
by the $T$--transforms. Thus it is a generalized cone bounded
by hyperplanes. This ``causal structure'' (to pursue the analogy
to relativity) induces a partial ordering of the states but there
is no complete ordering and in this sense then there does not exist
a preferred measure of entanglement, just as there does not exist
an absolute time in relativity. On the other hand (unlike in special
relativity) the underlying ``spacetime'' is not maximally symmetric,
although it comes close to be so. In such situations one often finds
that one time slicing is more ``natural'' than another. Pursuing the
analogy further one expects that a particular measure of entanglement
can be singled out by the physical context.

\bigskip

It is a pleasure to thank  Johan Br{\"a}nnlund,
Pawe{\l}~Horodecki, Marek Ku{\'s}
and Thomas Wellens
for inspiring discussions.
We appreciate helpful remarks raised by the referee.
Financial support by
Komitet Bada{\'n} Naukowych in Warsaw under
the grant 2P03B-072~19, the NFR,
and the European Science Foundation are gratefully acknowledged.

\appendix
\section{Distances in Weyl chambers as distances between orbits}

Before discussing possible interpretation of the distance in Weyl
chamber let us formulate two simple lemmas related to the majorization
theory.

{\bf Lemma 1}.
{\sl 
 Let $\vec x$, $\vec y$ and $\vec z$ be three
normalized probability
vectors  with $N$ components in the descending order each,
and let
$\vec{z}  \prec {\vec y}$. Then the scalar products fulfill
}
\begin{equation}
\vec{x} \cdot {\vec z} \le \vec{x} \cdot {\vec y}.
 \label{majorscla}
\end{equation}

{\bf Proof:} For simplicity we start with $N=3$ and consider the
difference $\vec{x} \cdot {\vec y} - \vec{x} \cdot {\vec z}$.
It is equal to the sum of three terms,
$x_1(y_1-z_1)+x_2(y_2-z_2)+x_3(y_3-z_3)=
 (x_1-x_2)(y_1-z_1)+(x_2-x_3)(y_1+y_2-z_1-z_2)
 +x_3(y_1+y_2+y_3-z_1-z_2-z_3)$, the first two are non-negative
and the last one vanishes. In the similar way, for an arbitrary $N$ we
represent $\sum_{k=1}^N x_k(y_k-z_k)$ as
$\sum_{k=1}^{N-1} (x_k-x_{k+1})\sum_{i=1}^k (y_i-z_i)$.
Due to ordering of the components of $\vec x$ and the majorization
relation $\vec{z}  \prec {\vec y}$ this expression
contains non-negative terms only, and this proves
the thesis (\ref{majorscla})  $\square$.
\medskip

Although the next lemma we shall use is closely related 
with recent results of Nielsen \cite{Ni99b}, 
we formulate it in the form most suitable for our purposes and 
for consistency provide its proof.

\smallskip 

{\bf Lemma 2}. 
{\sl Let $\rho$ be a density matrix, $\vec \lambda$ its
spectrum and ${\vec x}$ denotes the diagonal elements of $\rho$
represented in a certain basis. Then}
 $\vec{x}  \prec {\vec \lambda}$.

{\bf Proof:}
Any density matrix may be diagonalized, $\rho=U\lambda U^{\dagger}$.
Thus the diagonal elements  read
 $x_i=\rho_{ii}=U_{ij}\lambda_j U^{\dagger}_{ji}$. Introducing a
unistochastic matrix $D_{ij}:=|U_{ij}|^2$ we may then write
 ${\vec x}= D {\vec \lambda}$, so
 due to the Horn lemma  \cite{Ho54} the majorization relation
 ${\vec x}  \prec {\vec \lambda}$ holds  $\square $.

\medskip

Now we are ready to formulate statements concerning the Weyl chamber
which contains the set of diagonal density matrices with all entries
ordered.

{\bf Proposition 1:}
{\sl  Let $h$ and $g$ be two diagonal density
matrices of size $N$ with all components in a descending order, so
they belong to the same Weyl chamber.
Let $U$ and $V$ denote arbitary unitary matrices of size $N$.
Then

a) $D_{HS}(h,g) \le D_{HS}(UhU^{\dagger} , VgV^{\dagger})$,

b) the Euclidean distance between any two points in the Weyl chamber is
equal to the Hilbert--Schmidt distance between the global orbits
they generate.
}

\smallskip 

{\bf Proof:}. Let $W=U^{\dagger}V$.
Applying the definition $[D_{HS}(h,g)]^2={\rm Tr}(h-g)^2$
we may rewrite the difference of the squared distances,
 $Q=[D_{HS}(UhU^{\dagger} , VgV^{\dagger})]^2-
 [D_{HS}(h,g)]^2$  as $-2{\rm Tr}W^{\dagger} h W g
+2{\rm Tr} h g$.
Let ${\vec h}'$ denote the vector of the diagonal elements of
$W^{\dagger} h W$, so the difference we compute
is $Q=2({\vec h} \cdot {\vec g} -
     {\vec h}' \cdot {\vec g})$.
Applying lemma 2 we obtain ${\vec h}'\prec {\vec d}$,
and using lemma 1 we conclude that the difference $Q$ is non--negative,
which gives the statement a). Thus the HS--distance between two orbits,
$UhU^{\dagger}$ and $VgV^{\dagger}$, equal to the
HS--distance between closest points $h$ and $g$,
reads $D_{HS}(h,g)=[\sum_{i=1}^N (h_i-g_i)]^{1/2}$.
This is just the standard expression for the Euclidean distance between
two points in the Weyl chamber which represent both diagonal density
matrices, $h$ and $g$ $\square$. 

A similar statement may also be formulated for other distances
(trace,
Bures), but they do not allow for such a simple geometric interpretation
as does the Hilbert-Schmidt distance used above. Note that the assumption
that both density matrices belong to the same Weyl chamber is crucial,
since a generic orbit visits the simplex of eigenvalues $N!$ times,
corresponding to permutations of the components in the spectrum.

The above proposition clarifies the geometry of
Fig. 5a, which shows the asymmetric part of the simplex of eigenvalues
of the density matrices. To characterize in an analogous way the part of
the Schmidt simplex of pure states presented in Fig. 6a, we need to
change its geometry.
The equilateral triangle from Fig. 6b has to be deformed into
an octant of the sphere. More precisely, we take the vector of the
Schmidt coefficients
${\vec \lambda}=\{\lambda_1,...,\lambda_N\}$
of a pure state of a $N \times N$ bi-partite system and
transform it to $\{\lambda_1^2,...,\lambda_N^2 \}.$
Then the $N-1$ dimensional Schmidt simplex is mapped into
a $1/2^{N+1}$ part of the hypersphere $S^{N-1}$, equipped with a
natural Riemannian distance (for $N=3$ it is just the length of the arc
at the sphere $S^2$). This geometry allows us to formulate
an analogous

\smallskip

{\bf Proposition 2:} 
{\sl 
Let $\lambda$ and $\mu$ be two vectors of
Schmidt coefficients, representing two pure states $|\phi\rangle$
and $|\psi\rangle$  of a $N \times N$ bi--partite system.
Assume that all their components are ordered in a descending manner,
so both points belong to the same Weyl chamber.
Let $U=U_1\otimes U_2$ and $V=V_1\otimes V_2$ denote arbitary local
unitary transformations. Then

a) $D_{FS}(|\psi\rangle,|\phi\rangle) \le
    D_{FS}(U_1\otimes U_2 |\psi\rangle,V_1\otimes V_2|\phi\rangle)$,

b) the Riemannian distance between any two points in the Weyl
chamber (a subset of the deformed Schmidt simplex)
is equal to the Fubini--Study  distance between
the local orbits they generate.
}

\smallskip
{\bf Proof:}.
In the analogy to the proposition 1, the proof of this proposition may
be based on the lemmas 1 and 2. Alternatively, one can make use of
lemma 1 of a recent paper of Vidal et al. \cite{VJN00}, which states
that the maximal fidelity $F=|\langle \phi|U_1\otimes U_2
|\psi\rangle|^2$ is obtained for $U_1=U_2={\mathbb I}$ and is equal to
$F_{\rm max}=(\sum_{k=1}^N \sqrt{\lambda_k \mu_k})^2$.
This fact allows us to establish part b) of the proposition,
since the Riemannian distance between two points in the Weyl chamber
$D_R({\vec \lambda} , {\vec \mu})$ is equal to 
$\cos^{-1}(\sqrt{F_{\rm max}})=\frac{1}{2} \cos^{-1}(2{F_{\rm max}}-1)$
 and coincides with the Fubini--Study distance between
pure states $|\phi\rangle$ and $|\psi\rangle$ $\square$.

\smallskip

Note that the both propositions resemble the situation for a fibre bundle, 
where a metric on the total bundle space gives a shortest distance between 
the fibres that can then be used to define a metric on the base manifold. 
However, in our situation not all orbits need to be the same:
the dimensionality  of local orbits generated by
degenerated Schmidt vectors (located at the boundary of the Weyl chamber)
is analyzed in ref \cite{Si01}.

\end{document}